\documentclass[aps,prl,twocolumn,superscriptaddress,longbibliography]{revtex4-1}
\usepackage[utf8]{inputenc}
\usepackage{graphicx}
\usepackage{amsmath,amsfonts}
\usepackage[amssymb,Gray]{SIunits}
\usepackage{bm}
\usepackage{xcolor}

\renewcommand{\vec}[1]{\mathbf{#1}}

\newcommand{\com}{c.m.}

\newcommand{\prlsec}[1]{\textit{#1.---}}

\begin{document}


\title{Reply to Comment on ``Quantum mechanics for non-inertial observers''}


%
\author{Marko Toro\v{s}}
\email[]{marko.toros@ts.infn.it}
\affiliation{Department of Physics, University of Trieste, Strada Costiera 11, 34151 Miramare-Trieste, Italy}
\affiliation{Istituto Nazionale di Fisica Nucleare, Sezione di Trieste, Via Valerio 2, 34127 Trieste, Italy}
\author{Andr\'e Gro{\ss}ardt}
\email[]{andre.grossardt@ts.infn.it}
\affiliation{ZARM, University of Bremen, Am Fallturm 2, 28359 Bremen, Germany}
\author{Angelo Bassi}
\email[]{bassi@ts.infn.it}
\affiliation{Department of Physics, University of Trieste, Strada Costiera 11, 34151 Miramare-Trieste, Italy}
\affiliation{Istituto Nazionale di Fisica Nucleare, Sezione di Trieste, Via Valerio 2, 34127 Trieste, Italy}


\date{\today}
\begin{abstract}
Our recent paper (arXiv:1701.04298 {[}quant-ph{]}) discussed the occurrence
of a coupling of centre of mass and internal degrees of freedom for
complex quantum systems in non-inertial frames. There, we pointed
out that an external force supporting the system against gravity plays
a crucial role for the coupling between center of mass position to
the internal degrees of freedom. In a comment (arXiv:1702.06670 {[}quant-ph{]})
to our paper, Pikovski et~al.\ question our conclusion and present
the argument that the lack of the coupling term would be in contradiction
with the observation of gravitational time dilation using atomic clocks.
Here, we elaborate on our results in reply to their criticism and
clarify why our arguments remain valid.
\end{abstract}
\maketitle
Although a relativistic quantum mechanical description of a conserved
number of interacting particles does not exist~\cite{Currie:1963},
and the notion of centre of mass (\com) coordinates is ambiguous
in relativistic contexts~\cite{Pryce:1948}, a formalism introduced
by Krajcik and Foldy~\cite{Krajcik:1974} allows for a well-defined
discussion of relativistic corrections to the Schr\"odinger equation
in situations where particle creation and annihilation can be neglected.

Using this formalism, we have recently shown~\cite{Toros:2017} how
relativistic corrections to the dynamics of a many-body quantum systems
can be derived in non-inertial frames of reference: differently to previous more heuristic arguments,  our derivation is mathematically well grounded and
is based on the  map one can construct between symmetries and observables
in (symplectic) Hamiltonian mechanics~\cite{da2001lectures}. 

One outcome of the analysis is that the Hamiltonian of a composite  system of interacting quantum particles with total rest mass $M$, in a homogeneous gravitational potential, such as that of the Earth, takes the form:
\begin{equation}\label{eqn:hrindler}
\begin{split}
H &= H_\text{cm} + \left(1-\frac{\vec P^2}{2 M^2 c^2} + \frac{g\,X}{c^2} \right) H_\text{rel} + U_\text{ext} \,,
\end{split}
\end{equation}
where the explicit form of $H_\text{cm}$ and $H_\text{rel}$ up to order $1/c^2$ can be found in Ref.~\cite{Toros:2017} and is not relevant for the present discussion. This confirms previous results~\cite{Pikovski:2015}. 
The term of interest here is $g X H_{\text{\tiny rel}}/c^2$, which couples the c.m. vertical position $X$  to the internal Hamiltonian, inducing a decoherence effect in specific situations, as discussed in Ref.~\cite{Pikovski:2015}.
$U_\text{ext}$  is an external (supporting) potential. $H$ has to be understood as a function of the c.m. position ${\bf R}$ and momentum ${\bf P}$, and of the relative coordinates and momenta, which we collectively denote as $\rho$ and $\pi$, respectively: $H = H({\bf R}, {\bf P}, \rho, \pi)$. 

A second outcome of our analysis---very unsurprisingly in the light of  Relativity---is that physical predictions are observer dependent, and the decoherence effect appears and disappears, and in general changes, depending on which frame one looks at things from. (``Observer'' and ``reference frame'' are used as synonymous.)  In Ref.~\cite{Toros:2017} we considered four specific situations, where different things happen, depending on the relative state of motion between system and observer.

In Ref.~\cite{Pikovski:2017} the authors criticize our analysis in two main respects: i)~the fact that in some cases the decoherence effect cancels would be in contradiction with experimental evidence for time dilation (with atomic clocks); ii)~the role of an external potential supporting the system against gravity would not be analyzed correctly.  As a reply, we: i) make clear why our result is in no contradiction with the observation of time dilation between two (identical) atomic clocks; ii) analyze the role of the supporting potential and show that it also leads to a coupling between c.m. and internal motion at order $1/c^2$.

\prlsec{1. Time dilation between atomic clocks}
Take a Rindler observer 1 holding a clock, which is then at {\it rest} with respect to this observer's local reference frame, whose coordinates are labeled with the index ``1''. According to Eq.~(\ref{eqn:hrindler}),  
the clock does not exhibit a coupling of \com\ position and clock Hamiltonian ($H_\text{rel}$): in this case ${\bf R}_1 = 0$ and ${\bf P}_1 = 0$,  and then the  Hamiltonian becomes $H(0, 0, \rho_1, \pi_1) =  H_\text{rel}(\rho_1, \pi_1) + U_\text{ext}(\rho_1, \pi_1)$. 
Take a second observer 2 holding a second identical clock, which is  at {\it rest}  with respect to this second observer's local reference frame.  As before, the Hamiltonian in this new frame reduces to $H(0, 0, \rho_2, \pi_2) =  H_\text{rel}(\rho_2, \pi_2) + U_\text{ext}(\rho_2, \pi_2)$.

As the two Hamiltonians (specifically, the two internal ones) have the same functional dependence on the respective coordinates, the two clocks tick at the same rate as measured by the nearby observers. This is what one means with identical clocks.
Now, the two Hamiltonians belong to \emph{different} reference frames, describing the evolution
in the local time coordinate of \emph{that} respective frame. These time coordinates are identical to
the proper times of the \emph{respective} Rindler observers.  

Now, comparing the two clocks
amounts to a comparison of the proper times of the two observers' world lines, which yields exactly
the well-known classical gravitational time dilation, as  elucidated for example in Weinberg's book~\cite{weinberg1972gravitation}. (See Appendix A1 for a detailed derivation, with specific reference to atomic clocks.)

The coupling term of \com\ position and internal Hamiltonian for a \emph{single} clock, which seems to be the key element of the criticism of Pikovski et~al.~\cite{Pikovski:2017} to our work,
is by no means necessary in order to explain the experimentally observed time dilation between two
different clocks. It is also irrelevant for the explanation whether the considered clocks are
described quantum mechanically or classically.

As our results are in no contradiction with gravitational time dilation as measured with atomic clocks, the supposed mistake found by Pikovski et~al.~\cite{Pikovski:2017}, i.e. the way we consider the supporting potential,  looses its scope. However, it is interesting and relevant to further discuss the role of this potential.

\prlsec{2. Supporting potential} Here the authors of Ref.~\cite{Pikovski:2017}  are partly right. We did consider a special, yet at least theoretically important situation, where a system described by the Hamiltonian $H$, irrespective of its internal state of motion, is held against gravity by a fixed external supporting potential  $U_\text{ext}$. 
Such a potential is essentially a `counter'-gravitational potential, as, for instance, in the Newtonian case an electric potential $\phi_{el} = -g m X/q$ would cancel gravity for a particle of mass $m$ and charge $q$. This choice of potential does not invalidate one of the outcomes of our analysis, i.\,e. that the decoherence effect is an effect of the relative state of motion between system and observer, which the authors of Ref.~\cite{Pikovski:2017}  seem not to question. 

However, inspired by the remarks of Pikovski et~al.~\cite{Pikovski:2017}, we found it relevant to enter into the details of the supporting potential. Consider first the case they consider: a potential $U_\text{ext}(X)$, function only of the \com\ (vertical) coordinate $X$, having a local minimum capable of holding the system against gravity. To be more specific, we choose $U_\text{ext}(X) = \alpha X^2/2$ as an example of such a potential. Now, according to Hamilton's equations of motion the stationary solution is: $X = - (Mg + gH_\text{rel}(\rho, \pi)/c^2)/\alpha$. This simple result has two relevant consequences: i) First, if the internal state of the system changes (for example, by exchanging energy with the environment as in the physical situation considered in Ref.~\cite{Pikovski:2015}), then the c.m. starts moving. This is another way of saying that if the system's energy changes, it weights less or more and therefore its original motional state is not in equilibrium anymore. Accordingly, in the situation envisaged in Ref.~\cite{Pikovski:2017} the \com\ is not held fixed, in general. ii) If one insists in holding the c.m. fixed, also when the internal energy changes in time, then $U_\text{ext}(X)$ must depend on the internal energy as well, opposite to what is claimed in Ref.~\cite{Pikovski:2017}. (See Appendix A2 for further details.)
  
This brings us to consider a realistic potential  $U_\text{ext}$ depending on the position and momentum of each particle. If we re-write the potential in the c.m. and internal coordinates, up to order $1/c^2$, we find that the general form of such a relativistic potential is
\begin{equation}
U_\text{ext} = U_\text{ext}^{(0)}({\bf R}) + \frac{1}{c^2}U_\text{ext}^{(1)}({\bf R}, {\bf P}, \rho, \pi),
\end{equation}
(see Appendix A3 for details), where $U_\text{ext}^{(0)}({\bf R})$ is the nonrelativistic case considered in Ref.~\cite{Pikovski:2017} and the $1/c^2$ term $U_\text{ext}^{(1)}$ couples the \com\  position (and momentum) to the internal variables, as the term  $g X H_{\text{\tiny rel}}$ does. This means the following: an equilibrium solution mathematically exists, however in such a case the coupling between \com\ and internal degrees of freedom is given not only by the term $g X H_{\text{\tiny rel}}/c^2$ but also by the term $U_\text{ext}^{(1)}({\bf R}, {\bf P}, \rho, \pi)/c^2$. As such, in a situation similar to that considered in Ref.~\cite{Pikovski:2015} the external potential gives an additional contribution to the coupling of internal motion and \com, which can in principle dominate or (partially) cancel the gravitational coupling.

A note. Up to this point, our discussion has been completely classical, as were the critical arguments in
Ref.~\cite{Pikovski:2017}. In the quantum mechanical situation, the stationarity condition for the state of motion translates to the condition that the expectation value $\langle X \rangle$ is stationary,
which is what we considered in Ref.~\cite{Toros:2017}. 

Hence we come to the following conclusion: the natural state of a system in gravity is that of free fall. In such a case, according to the equivalence principle and according to what we discussed in Ref.~\cite{Toros:2017}, there is no decoherence effect unless the observer is {\it non}-inertial. Accordingly, the effect cannot be attributed to the system itself, but to the relative state of motion between system and observer. A decoherence effect can be attributed to the system when its motion deviates from free fall due to an external potential. However, it is the potential that generates the $1/c^2$ coupling between c.m. and internal motion.  The  coupling given by gravity, originating from the non-inertial motion of the observer,  adds to it. Anyhow, the observed final decoherence effect will still depend on the relative state of motion with respect to the observer. Ultimately, the decoherence effect is an effect of Special Relativity, not of General Relativity.

\begin{acknowledgments}
\prlsec{Acknowledgments}
The authors acknowledge funding and support from {\small{INFN}} and the University of Trieste (FRA 2016).
A.G. acknowledges funding from the German Research Foundation (DFG). We acknowledge useful discussions with D. Giulini and comments by A.~Deriglazov, L. Di\'{o}si, P.\,J. Felix, L. Lusanna, and G.~Torrieri.
\end{acknowledgments}

\appendix

\section*{Appendix}

\setcounter{equation}{0}
\renewcommand{\theequation}{A\arabic{equation}}

\subsection*{A1: Gravitational time dilation} 

Gravitational time dilation can be found in basically all books on General Relativity. Here we review how it works, to stress the points of interest with respect to the discussion in the main text.

We consider four observers: one Rindler (= at rest in gravity) observer with coordinates $x^{\mu'}=(ct',x')$, and a nearby Minkowski (= free fall) observer with coordinates $X^{\mu}=(cT,X)$, instantaneously at rest with respect to the Rindler observer; a second Rindler observer with coordinates $\tilde{x}^{\mu}=(c\tilde{t}',\tilde{x}')$, shifted by a quantity $b$ in the vertical direction with respect to the first Rindler observer (the coordinate time must also change, see Eqs.~\eqref{eq:t1}, \eqref{eq:t2}); a second Minkowski observer, with coordinates $\tilde{X}^{\mu}=(c\tilde{T},\tilde{X})$,  instantaneously at rest with respect to the second Rindler observer. The two Minkowski observers, by construction, change from time to time, but at each time they see each other instantaneously at rest, and are simply shifted with respect to each other by a quantity $b$ in the vertical direction.

The Minkowski coordinates $X^{\mu}=(cT,X)$ and the Rindler coordinates $x^{\mu'}=(ct',x')$ of the first set of observers are related as follows:

\begin{alignat}{1}
cT & =\left(x'+\frac{c^{2}}{g}\right)\sinh(\frac{gt'}{c}),\\
X & =\left(x'+\frac{c^{2}}{g}\right)\cosh(\frac{gt'}{c})-\frac{c^{2}}{g},
\end{alignat}
(see Eqs. (S1) and (S2) of~\cite{Toros:2017}, with $\bar{t}'=0$ for simplicity and without loss of
generality). The Rindler observer, located at $x'=0$, has
 proper time equal to the coordinate time, i.\,e. $d\tau=dt'$, and
is subject to proper acceleration $g$, i.\,e.
the usual Rindler observer.

The second Rindler observer is at point $x'=b$, with proper time $d\tilde{\tau}=(1+\frac{gb}{c^{2}})dt'$ and  is, as
we will see, subject to proper acceleration $\tilde{g}=g/(1+gb/c^{2})$.
The coordinate transformations between the two Rindler observers therefore are:
\begin{alignat}{1}
\widetilde{x}' & =x'-b,\label{eq:t1}\\
\widetilde{t}' & =\left(1+\frac{gb}{c^{2}}\right)t'.\label{eq:t2}
\end{alignat}
As a consistency check, a short calculation shows that:

\begin{alignat}{1}
c\tilde{T }& = cT = \left(\tilde{x}'+\frac{c^{2}}{\tilde{g}}\right)\sinh(\frac{\tilde{g}\tilde{t}'}{c}),\label{eq:t3}\\
\tilde{X} & = X - b = \left(\tilde{x}'+\frac{c^{2}}{\widetilde{g}}\right)\cosh(\frac{\tilde{g}\tilde{t}'}{c})-\frac{c^{2}}{\tilde{g}},\label{eq:t4}
\end{alignat}
which is the expected coordinate transformation between the second Rindler observer and the two Minkowski obersvers.

To compute time dilation between two identical clocks held by the two Rindler observers, we  follow the same calculation as in the supplementary sections S1 and S2 of~\cite{Toros:2017}. Specifically, we construct the  two Hamiltonians for the two  Rindler observers, by referring to the two instantaneous Minkowski observers, as amply discussed in Ref.~\cite{Toros:2017}. They have the same form in the respective coordinates; we write explicitly the first one (Eq.~(9) in Ref.~\cite{Toros:2017}):
\begin{equation}\label{s:rindlerH}
H^\text{Rindler}_\text{\com}=H^\text{Mink.}_\text{\com}+ \frac{g}{2 c^2} \{ X,H^\text{Mink.}_\text{\com} \}+U_{\text{ext}}\,,
\end{equation}
 where $H^\text{Mink.}_\text{\com}= \sqrt{\vec{P}^2 c^2 + H_\text{rel}^2} $, and we have added an external potential $U_{\text{ext}}$. For the second Rindler observer, all coordinates should be replaced with the ``tilde'' coordinates and $g$ with $\tilde{g}=g/(1+gb/c^{2})$. 
 
 We now Taylor expand the external potential in the center of mass position $X$:
\begin{equation}\label{s:external}
U_{\text{ext}}=U_{0} + U_{1} X + \frac{1}{2!} U_{2} X^2 + ...\,,
\end{equation}
where $U_{j}=\frac{\partial^j U_{\text{ext}}}{\partial X^j}|_{X=0}$, might still depend on the internal coordinates and c.m. momentum. Since the clock is held by the Rindler observer, i.\,e.  $\bm{R}=0$, $\bm{P}=0$ \footnote{A more general case $\bm{R}=0$, ${\bf P}=\text{const}\neq0$ does not change the argument.}, using Eqs.~\eqref{s:rindlerH}, \eqref{s:external} we obtain:
\begin{equation}\label{s:clock}
H^\text{Rindler}_\text{\com}= H_\text{rel}+U_{0}\,.
\end{equation}
The same is true for the second  clock, with respect to the second Rindler observer. In the case considered in Ref.~\cite{Toros:2017} (see Eq.~(12)), one immediately sees that $U_{0} = 0$.

Now we can address gravitational time dilation with atomic clocks~\cite{pound1959gravitational,Chou:2010}.
In this case the observed time dilation emerges by \emph{comparing} the two
\emph{identical clocks }(atoms) which are at \emph{different heights}. Since the comparison cannot be instantaneous, and since the Minkowski observers so far introduced change from time to time, we introduce a fifth fixed Minkowski observer with coordinates $x^{\mu}=(ct,x)$, who at time $t = 0$ is instantaneously at rest with all four observers so far introduced and at that time is located at $x' = 0$, and we describe the situation from her perspective.   

Suppose that at time $t = 0$, a photon is \emph{emitted} by the second clock, which is located at $x = b$; the photon encodes the information about the clock's ticking rate. The photon's frequency is then \emph{compared} with the ticking rate of the first clock, which by the time the photon reaches it, has been uniformly accelerated from the initial point $x = 0$ at $t=0$. 

At the time the photon is emitted, all five reference frames are at rest with respect to each other, therefore the photon's properties can be easily translated from one frame to any of the others. We first define them with respect to the second Rindler observer (Eq.~\eqref{s:clock}), holding the emitting clock: the energy is known (as given by the atomic transition, as measured by that observer), as well as its direction of motion (it must reach the first clock); it also follows a null geodesics: this fixes  the four-momentum. Then one can easily rewrite the four-vector in the Minkowski frame of the fifth observer: we call it $p^\mu$.  

\begin{figure}[h]
\includegraphics[width=8cm]{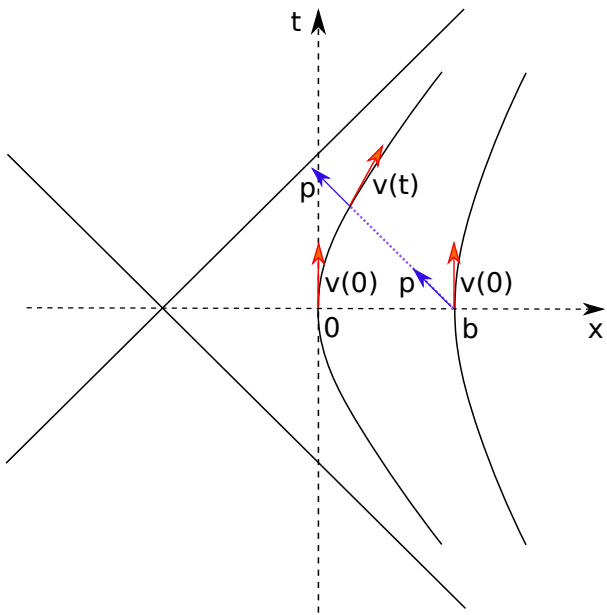}
\caption{Graphical depiction of gravitational time dilation (or redshift) from the perspective of the fifth Minkowski (inertial) observer. A photon  of four-momentum $p$ is emitted at $x = b$ at time $t=0$ and  at a later time $t>0$ is absorbed by a detector, which is uniformly accelerated from the initial point $x = 0$. At $t=0$ the four-velocity  of both the emitter atom and detector is $v(0)$, while at $t>0$ the four-velocity of the detector is $v(t)$. The energy of the photon as measured by the detector, is $E(t)=-\eta_{\mu\nu}p^\mu v^\nu (t)$.}
\label{Fig1}
\end{figure}

So far we considered the situation at time $t = 0$. Now we follow the motion of the photon, as described by the fifth inertial observer. This is easy: since  the motion is free, four-momentum is conserved. This is the advantage of describing the situation from the point of view of the fifth Minkowski observer.  Now the question is, what  is the photon's energy as measured by the first Rindler observer, or equivalently by the corresponding inertial observer instantaneously at rest, at the time the photon is absorbed. Since at that time these two observers are moving with velocity $v(t)$ as seen by the fifth Minkowski observer, Relativity tells that the energy they measure is~\cite{Misner:1973}: $E(t)=-\eta_{\mu\nu}p^{\mu}v^{\nu}(t)$,
where $v^{\nu}$ denotes the four-velocity of the Rindler observer, in the coordinates of the fifth Minkowski observer and $\eta_{\mu\nu}=\text{diag}(-1,+1,+1,+1)$. When the photon is absorbed by the first clock, its energy is shifted to~\cite{gourgoulhon2013special}:
\begin{equation}
E_\text{measured}=\left(1+\frac{gb}{c^{2}}\right)E_\text{emitted},
\end{equation}
which is the usual formula for time dilation. See Fig.~\ref{Fig1} for a representation of the whole emission/detection process. 

The calculation shows that time dilation is not related to the coupling between \com\ position ($X$) and the internal energy of a single clock ($H_{\text{rel}}^{(0)}$).

\subsection*{A2: Equilibrium points} We consider the Rindler Hamiltonian in Eq.~\eqref{s:rindlerH} and impose the condition of stationary \com, i.\,e.
\begin{equation}
\dot{\bm{R}}=\frac{\partial H^\text{Rindler}_\text{\com}}{\partial \bm{P}}=0 \,, \;\;\;\;\; \dot{\bm{P}}=-\frac{\partial H^\text{Rindler}_\text{\com}}{\partial \bm{R}}=0,
\end{equation}
where $U_{\text{ext}}(X)=\frac{\alpha X^2}{2}$ is a harmonic potential. It is straightforward to obtain the conditions:
\begin{equation}
\frac{\bm{P}}{H}=0\,, \;\;\;\;\; \
\frac{gH}{c^2}=-\frac{\partial U}{\partial X}. 
\end{equation}
The first one implies $\bm{P}=0$, while from the second, using  explicitly the harmonic form of the potential, we obtain $X=-\frac{g H}{\alpha c^2}$. In particular, expanding up to order $1/c^2$ we obtain:
\begin{equation}
X=-\frac{Mg + \frac{g H^{(0)}_\text{rel}}{c^2}}{\alpha },
\end{equation}
where $M$ is the total mass of the system.

\subsection*{A3: External potentials} 

Typical non-relativistic interaction potentials  depend only on the relative distance between the particles (Eq.~(3.1d) of Ref.~\cite{Krajcik:1974}). However, equally typical, relativistic
corrections, in the formalism of Krajcik
and Foldy, depend also on their momenta (Eq.~(3.1e)
of Ref.~\cite{Krajcik:1974}). Thus we assume that the external potential depends on both positions and momenta, i.\,e.:
\begin{equation}\label{s:potential}
 U_\text{ext}=\sum_{j=1}^N U(\bm{x}_j,\bm{p}_j),
\end{equation}
where $\bm{x}_j$, $\bm{p}_j$ denote the position and momentum of the $j$-th particle, with mass $m_j$. 

The expression for the \com\ coordinates in terms of the individual particles' coordinates are modified at order $1/c^2$ as follows (Eqs.~(2.27a),~(2.27b) of Ref.~\cite{Krajcik:1974}, respectively):
\begin{alignat}{1}
\bm{x}_{j} & =\bm{R}+\rho_{j}+\frac{1}{c^{2}}\chi_{j}(\bm{P},\rho,\pi),\label{eq:positions}\\
\bm{p}_{j} & =\frac{m_{j}}{M}{\bm{P}}+\pi_{j}+\frac{1}{c^{2}}\Pi_{j}(\bm{P},\pi),\label{eq:momenta}
\end{alignat}
where $\bm{R}$, $\bm{P}$ denote the \com\ postion and momentum, $\rho_j$, $\pi_j$  the relative position and momentum of the $j$-th particle,  $\rho$, $\pi$  collectively the relative positions and momenta, $\chi_{j}$, $\Pi_{j}$  the relativistic corrections  and $M$ is the total mass, i.\,e. $M=\sum_{j=1}^N m_j$. However, since $U_\text{ext}$ already depends at
order $1/c^{2}$ on momenta, we can neglect $\Pi$.
On the other hand, the correction $\chi$ is always present at order $1/c^2$: the relative momenta are an intrinsic part of any multiparticle potential at that order.

We now expand Eq.~\eqref{s:potential} up to order $1/c^2$:
\begin{equation}\label{s:electromagnetic}
U_\text{ext} = U_\text{ext}^{(0)}({\bf R},\rho) + \frac{1}{c^2}U_\text{ext}^{(1)}({\bf R}, {\bf P}, \rho, \pi),
\end{equation}
where $U_\text{ext}^{(0)}$, $U_\text{ext}^{(1)}$ denote the nonrelativistic and and first relativistic contribution, respectively. One normally assumes, when considering non-relativistic potentials, that over the volume of the system the potential is constant: this implies that $U_\text{ext}^{(0)}$ depends only weakly on the relative degrees of freedom $\rho$. In particular, by neglecting this dependence we obtain:
\begin{equation}\label{s:electromagnetic2}
U_\text{ext} = U_\text{ext}^{(0)}({\bf R}) + \frac{1}{c^2}U_\text{ext}^{(1)}({\bf R}, {\bf P}, \rho, \pi).
\end{equation}
On the other hand, for a generic internal state of motion, we cannot neglect the dependence of  $U_\text{ext}^{(1)}$ on the relative momenta $\pi$. This shows that a potential will in general couple in a complicated way the center of mass vertical position $X$ with the relative degrees of freedom.

%

\begin{thebibliography}{13}%
\makeatletter
\providecommand \@ifxundefined [1]{%
 \@ifx{#1\undefined}
}%
\providecommand \@ifnum [1]{%
 \ifnum #1\expandafter \@firstoftwo
 \else \expandafter \@secondoftwo
 \fi
}%
\providecommand \@ifx [1]{%
 \ifx #1\expandafter \@firstoftwo
 \else \expandafter \@secondoftwo
 \fi
}%
\providecommand \natexlab [1]{#1}%
\providecommand \enquote  [1]{``#1''}%
\providecommand \bibnamefont  [1]{#1}%
\providecommand \bibfnamefont [1]{#1}%
\providecommand \citenamefont [1]{#1}%
\providecommand \href@noop [0]{\@secondoftwo}%
\providecommand \href [0]{\begingroup \@sanitize@url \@href}%
\providecommand \@href[1]{\@@startlink{#1}\@@href}%
\providecommand \@@href[1]{\endgroup#1\@@endlink}%
\providecommand \@sanitize@url [0]{\catcode `\\12\catcode `\$12\catcode
  `\&12\catcode `\#12\catcode `\^12\catcode `\_12\catcode `\%12\relax}%
\providecommand \@@startlink[1]{}%
\providecommand \@@endlink[0]{}%
\providecommand \url  [0]{\begingroup\@sanitize@url \@url }%
\providecommand \@url [1]{\endgroup\@href {#1}{\urlprefix }}%
\providecommand \urlprefix  [0]{URL }%
\providecommand \Eprint [0]{\href }%
\providecommand \doibase [0]{http://dx.doi.org/}%
\providecommand \selectlanguage [0]{\@gobble}%
\providecommand \bibinfo  [0]{\@secondoftwo}%
\providecommand \bibfield  [0]{\@secondoftwo}%
\providecommand \translation [1]{[#1]}%
\providecommand \BibitemOpen [0]{}%
\providecommand \bibitemStop [0]{}%
\providecommand \bibitemNoStop [0]{.\EOS\space}%
\providecommand \EOS [0]{\spacefactor3000\relax}%
\providecommand \BibitemShut  [1]{\csname bibitem#1\endcsname}%
\let\auto@bib@innerbib\@empty
\bibitem [{\citenamefont {Currie}\ \emph {et~al.}(1963)\citenamefont {Currie},
  \citenamefont {Jordan},\ and\ \citenamefont {Sudarshan}}]{Currie:1963}%
  \BibitemOpen
  \bibfield  {author} {\bibinfo {author} {\bibfnamefont {D.~G.}\ \bibnamefont
  {Currie}}, \bibinfo {author} {\bibfnamefont {T.~F.}\ \bibnamefont {Jordan}},
  \ and\ \bibinfo {author} {\bibfnamefont {E.~C.~G.}\ \bibnamefont
  {Sudarshan}},\ }\bibfield  {title} {\enquote {\bibinfo {title} {Relativistic
  invariance and hamiltonian theories of interacting particles},}\ }\href@noop
  {} {\bibfield  {journal} {\bibinfo  {journal} {Rev. Mod. Phys.}\ }\textbf
  {\bibinfo {volume} {35}},\ \bibinfo {pages} {350--375} (\bibinfo {year}
  {1963})}\BibitemShut {NoStop}%
\bibitem [{\citenamefont {Pryce}(1948)}]{Pryce:1948}%
  \BibitemOpen
  \bibfield  {author} {\bibinfo {author} {\bibfnamefont {M.~H.~L.}\
  \bibnamefont {Pryce}},\ }\bibfield  {title} {\enquote {\bibinfo {title} {The
  mass-centre in the restricted theory of relativity and its connexion with the
  quantum theory of elementary particles},}\ }\href@noop {} {\bibfield
  {journal} {\bibinfo  {journal} {Proc. R. Soc. Lond. A}\ }\textbf {\bibinfo
  {volume} {195}},\ \bibinfo {pages} {62--81} (\bibinfo {year}
  {1948})}\BibitemShut {NoStop}%
\bibitem [{\citenamefont {Krajcik}\ and\ \citenamefont
  {Foldy}(1974)}]{Krajcik:1974}%
  \BibitemOpen
  \bibfield  {author} {\bibinfo {author} {\bibfnamefont {R.~A.}\ \bibnamefont
  {Krajcik}}\ and\ \bibinfo {author} {\bibfnamefont {L.~L.}\ \bibnamefont
  {Foldy}},\ }\bibfield  {title} {\enquote {\bibinfo {title} {Relativistic
  center-of-mass variables for composite systems with arbitrary internal
  interactions},}\ }\href@noop {} {\bibfield  {journal} {\bibinfo  {journal}
  {Phys. Rev. D}\ }\textbf {\bibinfo {volume} {10}},\ \bibinfo {pages}
  {1777--1795} (\bibinfo {year} {1974})}\BibitemShut {NoStop}%
\bibitem [{\citenamefont {Toro\v{s}}\ \emph {et~al.}(2017)\citenamefont
  {Toro\v{s}}, \citenamefont {Großardt},\ and\ \citenamefont
  {Bassi}}]{Toros:2017}%
  \BibitemOpen
  \bibfield  {author} {\bibinfo {author} {\bibfnamefont {Marko}\ \bibnamefont
  {Toro\v{s}}}, \bibinfo {author} {\bibfnamefont {André}\ \bibnamefont
  {Großardt}}, \ and\ \bibinfo {author} {\bibfnamefont {Angelo}\ \bibnamefont
  {Bassi}},\ }\href {https://arxiv.org/abs/1701.04298} {\enquote {\bibinfo
  {title} {Quantum mechanics for non-inertial observers},}\ } (\bibinfo {year}
  {2017}),\ \bibinfo {note} {arXiv:1701.04298 [quant-ph]},\ \Eprint
  {http://arxiv.org/abs/1701.04298} {arXiv:1701.04298 [quant-ph]} \BibitemShut
  {NoStop}%
\bibitem [{\citenamefont {Da~Silva}(2001)}]{da2001lectures}%
  \BibitemOpen
  \bibfield  {author} {\bibinfo {author} {\bibfnamefont {Ana~Cannas}\
  \bibnamefont {Da~Silva}},\ }\href@noop {} {\emph {\bibinfo {title} {Lectures
  on symplectic geometry}}},\ Vol.\ \bibinfo {volume} {1764}\ (\bibinfo
  {publisher} {Springer Science \& Business Media},\ \bibinfo {year}
  {2001})\BibitemShut {NoStop}%
\bibitem [{\citenamefont {Pikovski}\ \emph {et~al.}(2015)\citenamefont
  {Pikovski}, \citenamefont {Zych}, \citenamefont {Costa},\ and\ \citenamefont
  {Brukner}}]{Pikovski:2015}%
  \BibitemOpen
  \bibfield  {author} {\bibinfo {author} {\bibfnamefont {Igor}\ \bibnamefont
  {Pikovski}}, \bibinfo {author} {\bibfnamefont {Magdalena}\ \bibnamefont
  {Zych}}, \bibinfo {author} {\bibfnamefont {Fabio}\ \bibnamefont {Costa}}, \
  and\ \bibinfo {author} {\bibfnamefont {\u{C}aslav}\ \bibnamefont {Brukner}},\
  }\bibfield  {title} {\enquote {\bibinfo {title} {Universal decoherence due to
  gravitational time dilation},}\ }\href {\doibase 10.1038/nphys3366}
  {\bibfield  {journal} {\bibinfo  {journal} {Nat. Phys.}\ }\textbf {\bibinfo
  {volume} {11}},\ \bibinfo {pages} {668--672} (\bibinfo {year} {2015})},\
  \Eprint {http://arxiv.org/abs/1311.1095} {arXiv:1311.1095 [quant-ph]}
  \BibitemShut {NoStop}%
\bibitem [{\citenamefont {Pikovski}\ \emph {et~al.}(2017)\citenamefont
  {Pikovski}, \citenamefont {Zych}, \citenamefont {Costa},\ and\ \citenamefont
  {Časlav Brukner}}]{Pikovski:2017}%
  \BibitemOpen
  \bibfield  {author} {\bibinfo {author} {\bibfnamefont {Igor}\ \bibnamefont
  {Pikovski}}, \bibinfo {author} {\bibfnamefont {Magdalena}\ \bibnamefont
  {Zych}}, \bibinfo {author} {\bibfnamefont {Fabio}\ \bibnamefont {Costa}}, \
  and\ \bibinfo {author} {\bibnamefont {Časlav Brukner}},\ }\href@noop {}
  {\enquote {\bibinfo {title} {Comment on "quantum mechanics for non-inertial
  observers"},}\ } (\bibinfo {year} {2017}),\ \bibinfo {note} {arXiv:1702.06670
  [quant-ph]},\ \Eprint {http://arxiv.org/abs/1702.06670} {arXiv:1702.06670
  [quant-ph]} \BibitemShut {NoStop}%
\bibitem [{\citenamefont {Weinberg}(1972)}]{weinberg1972gravitation}%
  \BibitemOpen
  \bibfield  {author} {\bibinfo {author} {\bibfnamefont {Steven}\ \bibnamefont
  {Weinberg}},\ }\href@noop {} {\emph {\bibinfo {title} {Gravitation and
  cosmology: principles and applications of the general theory of
  relativity}}},\ Vol.~\bibinfo {volume} {67}\ (\bibinfo  {publisher} {Wiley
  New York},\ \bibinfo {year} {1972})\BibitemShut {NoStop}%
\bibitem [{Note1()}]{Note1}%
  \BibitemOpen
  \bibinfo {note} {A more general case $\protect \bm {R}=0$, ${\protect \bf
  P}=\protect \text {const}\not =0$ does not change the argument.}\BibitemShut
  {Stop}%
\bibitem [{\citenamefont {Pound}\ and\ \citenamefont
  {Rebka~Jr}(1959)}]{pound1959gravitational}%
  \BibitemOpen
  \bibfield  {author} {\bibinfo {author} {\bibfnamefont {Robert~V}\
  \bibnamefont {Pound}}\ and\ \bibinfo {author} {\bibfnamefont
  {GA}~\bibnamefont {Rebka~Jr}},\ }\bibfield  {title} {\enquote {\bibinfo
  {title} {Gravitational red-shift in nuclear resonance},}\ }\href@noop {}
  {\bibfield  {journal} {\bibinfo  {journal} {Physical Review Letters}\
  }\textbf {\bibinfo {volume} {3}},\ \bibinfo {pages} {439} (\bibinfo {year}
  {1959})}\BibitemShut {NoStop}%
\bibitem [{\citenamefont {Chou}\ \emph {et~al.}(2010)\citenamefont {Chou},
  \citenamefont {Hume}, \citenamefont {Rosenband},\ and\ \citenamefont
  {Wineland}}]{Chou:2010}%
  \BibitemOpen
  \bibfield  {author} {\bibinfo {author} {\bibfnamefont {C.~W.}\ \bibnamefont
  {Chou}}, \bibinfo {author} {\bibfnamefont {D.~B.}\ \bibnamefont {Hume}},
  \bibinfo {author} {\bibfnamefont {T.}~\bibnamefont {Rosenband}}, \ and\
  \bibinfo {author} {\bibfnamefont {D.~J.}\ \bibnamefont {Wineland}},\
  }\bibfield  {title} {\enquote {\bibinfo {title} {Optical clocks and
  relativity},}\ }\href@noop {} {\bibfield  {journal} {\bibinfo  {journal}
  {Science}\ }\textbf {\bibinfo {volume} {329}},\ \bibinfo {pages} {1630--1633}
  (\bibinfo {year} {2010})}\BibitemShut {NoStop}%
\bibitem [{\citenamefont {Misner}\ \emph {et~al.}(1973)\citenamefont {Misner},
  \citenamefont {Thorne},\ and\ \citenamefont {Wheeler}}]{Misner:1973}%
  \BibitemOpen
  \bibfield  {author} {\bibinfo {author} {\bibfnamefont {Charles~W.}\
  \bibnamefont {Misner}}, \bibinfo {author} {\bibfnamefont {Kip~S.}\
  \bibnamefont {Thorne}}, \ and\ \bibinfo {author} {\bibfnamefont
  {John~Archibald}\ \bibnamefont {Wheeler}},\ }\href@noop {} {\emph {\bibinfo
  {title} {Gravitation}}}\ (\bibinfo  {publisher} {W. H. Freeman and Company},\
  \bibinfo {address} {San Francisco},\ \bibinfo {year} {1973})\BibitemShut
  {NoStop}%
\bibitem [{\citenamefont {Gourgoulhon}(2013)}]{gourgoulhon2013special}%
  \BibitemOpen
  \bibfield  {author} {\bibinfo {author} {\bibfnamefont {{\'E}ric}\
  \bibnamefont {Gourgoulhon}},\ }\bibfield  {title} {\enquote {\bibinfo {title}
  {Special relativity in general frames},}\ }\href@noop {} {\bibfield
  {journal} {\bibinfo  {journal} {Special Relativigy in General Frames, by E.
  Gourgoulhon. Graduate Texts in Physics. ISBN 978-3-642-37275-9. Berlin:
  Springer-Verlag, 2013}\ } (\bibinfo {year} {2013})}\BibitemShut {NoStop}%
\end{thebibliography}

\end{document}